\documentclass[fleqn,twoside]{article}
\def\be{\begin{equation}}
\def\ee{\end{equation}}
\def\bi{\bibitem}
\begin{document}

\title{Intermediate inflation or late time acceleration?}
\author{Abhik Kumar Sanyal$^1$}\footnotetext[1]{Relativity and Cosmology
Research Centre,\\
\indent Department of Physics, Jadavpur University,\\ \indent Calcutta - 700032, India.\\
\indent Electronic address:abhikkumar@gmail.com;~~sanyal\_ak@yahoo.com\\
} \maketitle
\begin{center}
Dept. of Physics, Jangipur College, Murshidabad, \noindent
West Bengal, India - 742213. \\

\end{center}
\noindent
\begin{abstract}
The expansion rate of `Intermediate inflation' lies between the
exponential and power law expansion but corresponding accelerated
expansion does not start at the onset of cosmological evolution.
Present study of `Intermediate inflation' reveals that it admits
scaling solution and has got a natural exit form it at a later
epoch of cosmic evolution, leading to late time acceleration. The
corresponding scalar field responsible for such feature is also
found to behave as a tracker field for both gravity with canonical
and some non-canonical form of kinetic term. Thus the so called
Intermediate inflation should be considered as yet another dark
energy model, with asymptotic de-Sitter expansion.
\end{abstract}
\noindent
\section{Introduction}
It is now almost certain that the Universe contains $70\%$ of dark
energy, which is evolving slowly in such a manner that at present
the equation of state parameter, $w<-1/3$, or more precisely,
$w\approx-1$, so that the Universe is presently accelerating (see
eg. \cite{ms} for a recent review). $\Lambda$CDM-model is the
simplest one, that can explain the present observable features of
the Universe. However, in order to comply the vacuum energy
density, $\rho_{vac}\approx10^{74}GeV^4$ (calculated from quantum
field theory as, - $\rho_{vac}\approx\frac{m_{pl}^4}{16\pi^2}$)
with the critical density, $\rho_{\Lambda}\approx10^{-47}GeV^4$
(related to the cosmological constant as -
$\rho_{\Lambda}=\frac{\Lambda
 m_{pl}^2}{8\pi}$), it requires to set up yet another energy scale in
particle physics. This problem is known as the ` coincidence
problem' when stated as - ` why $\Lambda$ took 15 Billion years to
dominate over other kinds of matter present in the Universe'? A
scalar field with dynamical equation of state $w_{\phi}$, dubbed
as quintessence field \cite{cds} appears to get rid of this
problem, which during ` slow role' over the potential, acquires
negative pressure and finally acts as effective cosmological
constant $(\Lambda_{eff})$. Nevertheless, this quintessence field
requires to be fine tuned for the energy density of the scalar
field $(\rho_{\phi})$ or the corresponding effective cosmological
constant $(\Lambda_{eff})$, to be comparable with the present
energy density of the Universe. Tracker fields \cite{zs},
\cite{tr}, are introduced to overcome the fine tuning problem.
Tracker fields have attractor like solutions in the sense that a
wide range of initial conditions (viz., a wide range of initial
values of $\rho_{\phi}$) rapidly converge to a common cosmic
evolutionary track with $\rho_{\Lambda}$, and finally settles down
to the present observable Universe, with
$\rho_{\phi}\approx\rho_{\Lambda}$. Thus, tracker solutions avoid
both the coincidence problem and the fine tuning problem without
any need for defining a new energy scale.
\par
The important parameter required to check for the existence of the
tracker solutions is $\Gamma=\frac{V''(\phi)V(\phi)}{V'(\phi)^2}$,
$V(\phi)$ being the scalar potential. For quintessence, the
condition for the existence of the tracker solution with
$w_{\phi}<w_{B}$, (where $w_{\phi}$ and $w_{B}$ are the state
parameters of the scalar field and the background field
respectively) is $\Gamma>1$, or equivalently, $|\lambda|=
|\frac{V'(\phi)}{\kappa V(\phi)}|\approx |\frac{H}{\dot \phi}|$
decreasing as $V(\phi)$ decreases. Tacker solution further
requires a nearly constant $\Gamma$, which is satisfied if
$|\frac{d(\Gamma-1)}{Hdt}|\ll|\Gamma-1|$, or equivalently,
$|\Gamma^{-1}\frac{d(\Gamma-1)}{Hdt}|\approx
|\frac{\Gamma'}{\Gamma(\frac{V'}{V})}|\ll1$ \cite{zs}. The
condition $w_{\phi}<w_{B}$ is required for the present day
acceleration of the Universe. So eventually the slope of the
potential becomes sufficiently flat ensuring accelerated expansion
at late times. The same condition for k-essence models, having
non-canonical form of kinetic energy requires $\Gamma>3/2$, and is
slowly varying \cite{tc}. Such a scalar field remains subdominant
until recently.
\par
General theory of relativity with a minimally coupled scalar field
admits a solution in the form $a = a_{0}~\exp{(At^f)}$, (where,
$a$ is the scale factor and $a_{0}>0,~ A>0$ and $0<f<1$ are
constants) which was dubbed as intermediate inflation in the
nineties \cite{b}, \cite{m}. The expansion rate in the
intermediate inflation is faster than power law and slower than
exponential ones. Some aspects of intermediate inflation have been
studied in the past years \cite{if}, \cite{rn}. Particularly, in a
recent work \cite{iff} it has been shown that such inflationary
model can encounter the observational features of the three year
Wilkinson Microwave Anisotropy Probe (WMAP) data \cite{wmap} with
spectral index $n_{s}=1$, considering non-zero tensor-to-scalar
ratio $r$. Such solutions \cite{b}, \cite{m} may also appear in
other theories of gravitation. In fact, the action obtained under
modification of Einstein's theory by the introduction of higher
order curvature invariant terms (which is essentially four
dimensional effective action of higher dimensional string
theories), has been found to be reasonably good candidate to
explain the presently observed cosmological phenomena. In
particular, in a recent work it has been found \cite{a} that
Gauss-Bonnet interaction in four dimensions with dynamic dilatonic
scalar coupling leads to a solution $a = a_{0}~\exp{(At^f)}$ in
the above form, where the Universe starts evolving with a
decelerated exponential expansion. Such solutions encompasses the
cosmological evolution, as the dilatonic scalar during evolution
behaves as stiff fluid, radiation and pressureless dust. Solutions
of these type are known as scaling solutions \cite{clw}, \cite{t}
in which the energy density of the scalar field $(\rho_{\phi})$
mimics the background matter energy density. It then comes out of
the scaling regime \cite{clw}, \cite{t} and eventually the
Universe starts accelerating. Asymptotically, the scalar behaves
as effective cosmological constant. The deceleration parameter
corresponding to such solution is given by
$q=-1+\frac{1-f}{Aft^f}$. Thus, unlike usual inflationary models
with exponential or power law expansion, accelerated expansion of
the scale factor corresponding to intermediate inflation \cite{b},
\cite{m} does not start at the onset of the cosmological
evolution, rather it starts after the lapse of quite some time.
\par
Inflation should have started at the Planck epoch so that it can
solve the initial conditions viz., the horizon and the flatness
problems of the standard model and can lead to almost a scale
invariant spectrum of density perturbation. As such the epoch at
which accelerated expansion of the scale factor in intermediate
inflation starts, has also been arbitrarily taken as the Planck's
era. But it is not true. Because, as observed in the context of
Gauss-Bonnet gravity \cite{a}, such solutions admit synchronize
scaling between $\rho_{\phi}$ and $\rho_{B}$, which can happen
long after the Planck's era. Thus it is required to study the so
called intermediate inflation in some more detail.
\par
A comprehensive study in the present work reveals that, (1)
solutions in the form $a = a_{0}~\exp{(At^f)}$, lead to late time
acceleration and therefore should be treated as dark energy model
rather than inflation. The nature of such solution also reveals
that the equation of state of the scalar field ($w_{\phi}$)
follows that of the background matter ($w_{B}$) closely, and
finally the scalar field comes out of the scaling regime
\cite{clw}, \cite{t}, leading to accelerated expansion of the
Universe. The scalar field also admits tracking condition
$\Gamma>1$, for standard form of kinetic energy. (2) Next, it has
been observed that even for a non-canonical form of kinetic energy
the same result is reproduced with the same form of potential,
which also satisfies the tracker condition $\Gamma>3/2$. (3) We
then proceed to show that such solutions with non-canonical form
of kinetic energy does not always carry a tracker field. (4)
Finally, it has been shown that in the presence of background
matter such solutions are admissible with a tracking field.
\par
In the following section, we have started with a k-essence action
\cite{ke} in it's simplest form, keeping only a coupling parameter
$g(\phi)$ in the kinetic energy term and write down the field
equations. In section 3, instead of choosing the form of the
potential, we have chosen different forms of the super-potential
$H(\phi)$ \cite{sp}, and presented explicit solutions in the form
$a = a_{0}~\exp{(At^f)}$, discussed above, for standard $g=1/2$
and nonstandard form of kinetic energy $g=g(\phi)$. Finally in
section 4, similar solutions in the presence of background matter
have been presented.

\section{Action and the field equations}
The generalized k-essence \cite{ke} non-canonical Lagrangian,
 \[ L = g(\phi)F(X) - V(\phi),\]
where, $X = \frac{1}{2} \partial_{\mu}\phi \partial^{\mu}\phi$,
when coupled to gravity may be expressed in the following most
simplest form

\be S=\int
d^4x\sqrt{-g}[\frac{R}{2\kappa^2}-g(\phi)\phi,_{\mu}\phi^{,~\mu}-V(\phi)],
\ee where, a coupling parameter, $g(\phi)$ is coupled with the
kinetic energy term. $g(\phi)$ has got a Brans-Dicke origin,
$g=\frac{\omega}{\phi}$ too, $\omega(\phi)$ being the Brans-Dicke
parameter. This is the simplest form of an action in which both
canonical and non-canonical forms of kinetic energies can be
treated. For the spatially flat Robertson-Walker space-time

\[ds^2=-dt^2+a^2(t)[dr^2+r^2\{ d\theta^2+ sin^2(\theta)d\phi^2\}]\]
the field equations are

\be 2\frac{\ddot a}{a}+\frac{\dot a^2}{a^2}=2\dot
H+3H^2=-\kappa^2[g\dot\phi^2-V(\phi)]=-8\pi G p, \ee and

\be 3\frac{\dot a^2}{a^2}=3H^2=\kappa^2[g\dot \phi^2+V(\phi)]=8\pi
G \rho , \ee where, $H=\dot a/a$, is the Hubble parameter. In
addition, we have got the $\phi$ variation equation

\be \ddot\phi+3\frac{\dot
a}{a}\dot\phi+\frac{1}{2}\frac{g'}{g}\dot\phi^2+\frac{V'}{2g}=0,
\ee which is not an independent equation, rather it is derivable
from the above two equations (2) and (3). In the above, over-dot
and dash $(')$ stand for differentiations with respect to time and
$\phi$ respectively. Instead of using equations (2) and (3), it is
always useful to parametrize the motion in terms of the field
variable $\phi$ \cite{m}, \cite{sbl}. Thus, with $\kappa^2=8\pi
G$, the above set of equations can be expressed as,

\be \dot\phi=-\left(\frac{H'(\phi)}{\kappa^2 g(\phi)}\right),\ee
and in the Hamilton-Jacobi form

\be[H'(\phi)]^2-3\kappa^2 H^2(\phi)g(\phi)+\kappa^4
V(\phi)g(\phi)=0.\ee The two important parameters of the theory
viz., the equation of state $w_{\phi}$ and the deceleration $q$
parameters are expressed as,

\be w_{\phi}=-1-\frac{2\dot H}{3H^2};\;\;q=-1-\frac{\dot H}{H^2}.
\ee Now, we are to solve for $a$ (in view of $H$), $\phi$,
$g(\phi)$, and $V(\phi)$ from the above two field equations (5)
and (6), and so, two additional assumptions are required. It is
found that some specific forms of the super-potential $H(\phi)$
\cite{sp} lead to the so called intermediate inflationary
solutions.

\section{Solution in the form $a = a_{0}e^{(At^f)}$ and it's dynamics}

This section is devoted in presenting the solution of the scale
factor in the form $a = a_{0}e^{(At^f)}$, which was dubbed as
intermediate inflation earlier. In order to study the dynamics of
such solution in detail, we have tacitly assumed the presence of
any form of background matter in this section.

\subsection{Case-I}
Let us choose the following form of the super-potential,
$H(\phi(t))$ as,

\be H=\frac{h}{\phi^n}, \ee where, $h>0$ and $n>0$ are constants.
In view of this assumption equation (5) becomes,

\be g\dot \phi=\frac{nh}{\kappa^2}~\phi^{-(n+1)}.\ee We can now
solve the set of equations (6), (8) and (9), provided we choose a
particular form of $g(\phi)$ or $V(\phi)$. In the following, we
choose the standard form of kinetic energy ie., $g=1/2$. For the
most natural choice $g=1/2$, the field variables are found from
equations (9) and (8) as,

\be \phi=\left[\frac{2hn(n+2)}{\kappa^2}
t\right]^{\frac{1}{n+2}};\;\;H=mt^{-(\frac{n}{n+2})}
   ;\;\;
a=a_{0}~\exp{\left[\left({\frac{m(n+2)}{2}}\right)
t^{(\frac{2}{n+2})}\right]},\ee where,
$m=h^{(\frac{2}{n+2})}\left[\frac{\kappa^2}{2n(n+2)}
\right]^{(\frac{n}{n+2})}$. The above form of the scale factor,
the Hubble parameter and the scalar field $\phi$ can be expressed
respectively as,

\be a = a_{0}\exp{[At^f]},\;\;\; \;\;H=\frac{A
f}{t^{(1-f)}},\;\;\;and \;\;\; \phi=\left[\frac{8h(1-f)}{\kappa^2
f^2}~t\right]^{\frac{f}{2}},\ee where, $A=\frac{m(n+2)}{2}>0$ and
$0<f=\frac{2}{n+2} <1$, are related to the constants $h$ and $n$.
In view of equation (7), the state parameter and the deceleration
parameter evolve as,

\be w_{\phi}=-1+\frac{n}{3A} t^{-f};\;\;q=-1+\frac{n}{2A}
t^{-f}.\ee The form of the potential, in view of equation (6), for
such a solution is restricted to,

\be V(\phi)=\frac{h^2}{\kappa^4}\left(\frac{3\kappa^2}{\phi^{2n}}
-\frac{2n^2}{\phi^{2(n+1)}}\right), \ee which has the form of
double inverse power. This solution was obtained by Barrow
\cite{b} and Muslimov \cite{m} in the nineties and was dubbed as
intermediate inflation, since the expansion rate of the scale
factor is greater than the power law but less than standard
exponential law. For such an expansion rate,

\be \rho+p=\frac{4n^2h^2}{\kappa^4\phi^{2(n+1)}}>0,\ee ie., the
weak energy condition is always satisfied, and so $w_{\phi}\geq
-1$. However,

\be \rho+3p=\frac{6}{\kappa^2}(-\dot
H-H^2)=\frac{6Af}{\kappa^2}[(1-f)-Aft^f]t^{-(2-f)},\ee implies
that the strong energy condition is violated at

\be t>\left(\frac{1-f}{A f}\right)^{\frac{1}{f}},\ee when
$w_{\phi} < -1/3$.

It is to be noted that the necessary condition for inflation,
viz., $\ddot a>0$ or more precisely,
$\frac{d}{dt}{(\frac{H^{-1}}{a})}<0$, is satisfied under the same
above condition (16). Eventually, for large $\phi$, viz.,
$\phi\gg\frac{\sqrt 2 n}{\kappa}$ the potential energy starts
dominating over the kinetic energy, ie., $\dot\phi^2\ll V(\phi)$,
and thus the slow roll condition, $\epsilon = -\frac{\dot H}{H^2}
<1$, is also satisfied under condition (16). However, Inflation is
supposed to have started at Planck's era, so that it can solve the
initial problems of the standard model, viz., the horizon and the
flatness problems, the structure formation problem and lead nearly
to a scale invariant spectrum. Here we observe that, accelerated
expansion of the scale factor in the so called intermediate
inflation does not start at the onset of cosmic evolution. So, now
we can ask the question, `what happened prior to the onset of the
accelerated expansion in the so called intermediate inflationary
era?'.
\par
The solution dictates that the Universe starts evolving from an
infinitely decelerated exponential expansion with $w_{\phi}>1$. It
might appear that we are considering  a highly unorthodox
cosmological model involving a ultra-hard equation of state and
super-luminal speed of sound. This is true in some sense, since as
mentioned earlier, in order to study the situations under which
such solutions emerge and it's dynamics, we have not considered
the presence of any form of background matter explicitly.
Nevertheless, this situation is quite similar to the phantom
models where, super negative pressure gives rise to ultra negative
equation of state, indicating that the effective velocity of sound
in the medium, $v=\sqrt{|\frac{dp}{d\rho}|}$ might become larger
than the velocity of light. Likewise, here the Universe starts
evolving with such a situation which actually demonstrates that
corresponding era is classically forbidden and the need for
invoking quantum cosmology at that era.
\par
Now, during evolution the Universe passes through the stiff fluid
era $w_{\phi}=1$, the radiation dominated era $w_{\phi}=1/3$, the
pressureless dust era $w_{\phi}=0$, the transition (from
deceleration to acceleration) era $w_{\phi}=-1/3$, and
asymptotically tends to the magic line, ie the vacuum energy
dominated inflationary era $w_{\phi}\approx-1$. Thus, the equation
of state $w_{\phi}$ follows the matter equation of state $w_{B}$
closely (which as already mentioned has been assumed in this
section tacitly) and so it corresponds to the scaling solution
\cite{clw}, \cite{t}. It finally comes out of the scaling regime
and enters into the transition era.
\par
Now, in the context of a realistic model in the presence of some
form of background matter, if the above solution (11) is
accompanied with a potential (13) then another important aspect
has to be checked, and that is if the solution is tracking, which
is true when the following condition is satisfied,
\[\Gamma=\frac{V''V}{V'^2}>1,\]
which is equivalent to check if
\[|\lambda|=|\frac{V'(\phi)}{\kappa V(\phi)}|\]
decreases with $V(\phi)$. Now, for the above form of the potential
(13) the tracking condition $\Gamma>1$ corresponds to
\[
9\kappa^4\phi^4-6n(2n+3)\kappa^2\phi^2+4n^3(n+1)>0,\] ie., the
solution starts tracking at the epoch when
\be\phi^2>\frac{n(2n+3)-n\sqrt{8n+9}}{3\kappa^2},\;\;\;ie.,\;\;
t>\frac{2(1-f)}{3\kappa^2f^2}[4-f-\sqrt{16f-7f^2}].\ee

Further,
\[|\frac{V'}{V}|=
2n\left(\frac{3\kappa^2\phi^2-2n(n+1)}{3\kappa^2\phi^2-2n^2}\right)
(\frac{1}{\phi}),\]
decreases with $V$ for large value of $\phi$, and asymptotically
vanishes, ie., for $\phi\rightarrow \infty$,
$|\frac{V'}{V}|=\frac{2n}{\phi}\rightarrow 0$. The other condition
that $\Gamma$ should be slowly varying is satisfied provided,
\[|\frac{\Gamma'}{\Gamma(\frac{V'}{V})}|<< 1.\] This condition
yields a ratio of two polynomials in $\phi$. The highest degree in
the numerator is $8$ while that in the denominator is $10$. So the
above condition is satisfied as $\phi$ evolves, since $\phi$ is a
monotonically increasing function of $t$. In particular, this
condition is satisfied for
\[\phi^2\gg\frac{2[9n(10n^2+15n+6)+1]}{27\kappa^2(2n+1)},\]
which corresponds to the epoch when the scalar field takes over
the matter dominated era, ie., $w_{\phi}<w_{B}$.

So the corresponding scalar field is a tracker field, which at the
end rolls down a sufficiently flat potential, since,
$|\lambda|=|\frac{V'}{\kappa V}|$ decreases with $V$ and finally
tends to zero. Thus both the coincidence and the fine tuning
problems are solved. Hence, the so called intermediate inflation
does not inflate the Universe at an early epoch, rather it leads
to the presently observable cosmic acceleration which avoids both
the coincidence and the fine tuning problem of earlier
quintessence models. So the intermediate inflation should be
treated as yet another dark energy model, and this proves our
first claim.
\subsection{Case-II}
In this subsection we show that the same set of solutions obtained
in case-I can be reproduced even for a non-canonical kinetic term.
To show this, we observe that the above set of solutions does not
necessarily require to fix up $g=1/2$, rather they are found even
for a functional form of $g=g(\phi)$, ie. with a non-canonical
kinetic term. This can be checked easily by assuming the solution
(11) of the scale factor along with the assumption (8) for the
Hubble parameter. The field $\phi$, the potential $V(\phi)$, and
the coupling parameter $g(\phi)$ are then found as,

\be
\phi=\left(\frac{h}{Af}\right)^{\frac{1}{n}}t^{(\frac{1-f}{n})};\;\;
V=\frac{h^2}{\kappa^2}\left[\frac{3}{\phi^{2n}}-
\left(\frac{h^f}{Af}\right)^{\frac{1}{1-f}}
\frac{(1-f)}{\phi^{n(\frac{2-f}{1-f})}}\right];\;\;
g=\frac{hn^2}{\kappa^2(1-f)}
\left(\frac{Af}{h}\right)^{\frac{1}{1-f}}\phi^{[\frac{nf}{1-f}-2]}.\ee
Since $(\frac{2-f}{1-f})>2$, therefore the potential is in the
same above form (13). To check if the potential is a tracker
field, we have to consider the corresponding condition for
k-essence models, given in \cite{tc}, viz.,

\[\Gamma=\frac{V''V}{V'^2}>\frac{3}{2}.\]
So let us express the potential given in (18) in the following
form,

\[V=\frac{A}{\phi^{N}}-\frac{B}{\phi^{M}},\]
where, $M(=n\frac{2-f}{1-f})>N(=2n)$, while the constants,
$A=\frac{3h^2}{\kappa^2}$ and
$B=\frac{1-f}{\kappa^2}(\frac{h^{(2-f)}}{Af})^{\frac{1}{1-f}}$.
Now, we observe that $\Gamma>3/2$ is satisfied provided,

\[A^2N(2-N)\phi^{2M}+B^2M(2-M)\phi^{2N}-2AB(M^2+N^2+M+N-MN)\phi^{M+N}>0,\]
which is true for $N<2$, ie., $n<1$, since as mentioned $M>N$ and
$\phi$ is a monotonically increasing function of time. The other
condition that $\Gamma$ is slowly varying, is also satisfied as
before, since, $|\frac{\Gamma'}{\Gamma(\frac{V'}{V})}|$ is found
as the ratio of two polynomials in $\phi$, with a lower highest
degree in the numerator than the denominator. Thus this solution
has got all the features, including tracking behaviour for
$w_{\phi}<w_{B}$, of the previous solution.
\par
Thus we find that even a non-canonical kinetic term reproduces the
same set of solutions obtained with a canonical kinetic term. So,
in principle it is possible to find a field theory with a
non-canonical kinetic term, such that the cosmological solution is
exactly the same as the solution of the said theory with canonical
kinetic term. This proves our second claim which is of-course a
new result.
\par
It is to be mentioned that for $f=\frac{2}{n+2}$, the
non-canonical kinetic term turns out to be canonical and the
situation arrived at in case-I is recovered.

\subsection{Case-III} It should not be taken as granted that solution
in the form $a = a_{0}e^{(At^f)}$ is always accompanied by a
potential which is a tracker field. In this subsection we show
that even a different form of the super-potential in the presence
of a nonstandard form of kinetic energy, leads to similar form of
the scale factor obtained in the previous subsections but it
carries with it a potential which does not satisfy tracking
condition. Let us choose the form of the super-potential
$H(\phi(t))$ as,

\be H=\kappa^2 e^{-l\phi}, \ee with $l>0$. So,

\be g\dot\phi=le^{-l\phi};\;\;\frac{\dot
H}{H^2}=-\frac{l^2}{\kappa^2
g};\;\;V=-le^{-l\phi}\dot\phi+3\kappa^2e^{-2l\phi}.\ee In view of
equation (7) and (19), it is clear that for a constant $g$,
$w_{\phi}$ becomes non-dynamical. So, to find the solutions
explicitly, let us further make the following choice,

\be\dot\phi=e^{-ml\phi},\ee where $m$ is a constant. Under this
choice,

\be g= le^{(m-1)l\phi},\ee and the potential can be expressed as
the algebraic sum of two inverse exponents as,

\be V(\phi)=3\kappa^2 e^{-2l\phi}-l e^{-(m+1)l\phi}.\ee This form
of the potential \cite{t}, \cite{as}, arise as a result of
compactifications in superstring models and is usually considered
to exit from the scaling regime. Solutions for the scalar field
and the scale factor are obtained as,

\be \phi=\frac{1}{ml}\ln(mlt);\;\; a=
a_{0}\exp[\frac{m\kappa^2}{(m-1)(ml)^\frac{1}{m}}t^\frac{m-1}{m}].\ee
Thus the scale factor can be expressed in the same form (11) of
the so called intermediate inflation for $m>1$. Further, the
equation of state and the deceleration parameters (7) are obtained
as,

\be w_{\phi}=-1+\frac{2l}{3\kappa^2}(mlt)^{\frac{1-m}{m}};\;\;
q=-1+\frac{l}{\kappa^2}(mlt)^{\frac{1-m}{m}}.\ee  Now, for $m>1$,
\be \rho+p=2l \exp[-(m-1)l\phi]>0,\ee ie., the weak energy
condition is always satisfied while, \be \rho+3p=6\left(l
\exp[-(m-1)l\phi]-\kappa^2\exp[-2l\phi]\right),\ee ie., the strong
energy condition is violated at $t \geq
\frac{1}{ml}(\frac{l}{\kappa^2})^{\frac{m}{m-1}}$, which
corresponds to $\exp{(m-1)l\phi}\geq \frac{l}{\kappa^2}$, - the
epoch of transition from decelerating to the accelerating phase.
Further, the necessary condition for inflation and the slow roll
condition are satisfied at the same epoch. The Universe expands
exponentially but decelerates from a infinitely large value. The
equation of state $w_{\phi}$ starts from indefinitely large value
at the beginning. $w_{\phi}\rightarrow\infty$, implies a greater
effective velocity of sound than that of light in the
corresponding medium, which also appears in phantom models having
super negative pressure. So classically it has got no meaning at
all. Such result only dictates the importance of invoking quantum
cosmology before the equation of state reaches stiff fluid era.

\par
As before, the field satisfies scaling solution, since during
evolution it takes over the phases of stiff fluid, radiation and
the matter dominated era. It then finally comes out of the scaling
regime and enters into an accelerating phase before asymptotically
it reaches the desired value of minus one. So this solution also
has got the same feature as the previous one. However,
\[\Gamma=\frac{V''(\phi)V(\phi)}{V'(\phi)}>3/2,\] requires
\[36\kappa^4e^{2ml\phi}+6\kappa^2l[(m+1)^2-2]
e^{(m+1)l\phi}+l^2(m+1)^2e^{2l\phi}<0,\] which is not satisfied
for $m>1$. So the potential (23) does not satisfy the tracker
condition. Thus fine tuning problem can not be avoided in this
model. Note that for $m<1$, the above relation may be satisfied,
but then the solution does not correspond to late time
acceleration, rather it is inflationary, in which the scale factor
starts from zero and ends up with the value $a_{0}$. Hence, our
third assertion that in non-canonical theories, the solution in
the form $a = a_{0}e^{(At^f)}$ does not always carry with it a
tracker field, has been established.

\section{Presence of background matter}
Observations suggest that our Universe is presently filled with
$70\%$ of dark energy, $26\%$ of dark matter, $4\%$ of Baryons and
$0.005\%$ of radiation \cite{tp}. So, to consider a realistic
model, presence of background distribution of all types of
Baryonic and non-Baryonic matter should be accounted for
explicitly. In this section our motivation is to check if the
above form of the scale factor admits viable cosmological solution
in the presence of background matter. The field equations now can
be arranged as,

\be \dot{H}=-\kappa^2[g\dot\phi^2+\frac{\rho_{B}+p_{B}}{2}],\ee

\be \dot{H}+3H^2=\kappa^2[V(\phi)+ \frac{\rho_{B}-p_{B}}{2}],\ee
where $\rho_{B}$ and $p_{B}$ are the energy
density and pressure of the background matter respectively. Further, since scalar field is minimally coupled to
the background, so continuity equations for the background matter and the scalar field hold independently.
Hence, We can write,

\be
\dot\rho_{B}+3H(1+w_{B})\rho_{B}=0=\dot\rho_{\phi}+3H(1+w_{\phi})\rho_{\phi}.\ee
Thus we have

\be \rho_{B}={\rho_{B}}^{(0)}~ a^{-3(1+w_{B})},\ee and

\be \rho_{\phi}={\rho_{\phi}}^{(0)}~ a^{-3(1+w_{\phi})},\ee where,
${\rho_{B}}^{(0)}$ and ${\rho_{\phi}}^{(0)}$ are the present
values of the background and the scalar field energy densities. If
we now plug in the solution of the scale factor in the form
$a=a_{0}\exp{At^f}$, with $a_{0}>0, A>0$ and $0<f<1$, then the
potential is found in view of equation (29) as,

\be V=\frac{1}{\kappa^2}[\frac{3A^2 f^2}{ t^{2(1-f)}}-\frac{A f(1-f)}{
t^{(2-f)}}]-\frac{(1-w_{B})\rho_{B}}{2},\ee
while the kinetic term can be obtained in view of equation (28) as,

\be g\dot\phi^2=\frac{A f(1-f)}{\kappa^2 t^{(2-f)}}-\frac{(1+w_{B})\rho_{B}}{2}.\ee In the above two equations
(33) and (34), we have used the equation of state $p_{B}=w_{B}\rho_{B}$, for the background matter. Now, it is
not possible to find a solution of $\phi$ in closed form for the standard form of kinetic energy ($g=1/2$). Thus
one can make some suitable choice of the scalar field $\phi$ to express the potential $V(\phi)$ and the coupling
parameter $g(\phi)$ as functions of $\phi$. It is also noticed that the first term of the potential is
predominantly dominating as the Universe evolves. Thus the potential remains positive asymptotically, though it
starts from an indefinitely large negative value. As an example, let us choose $\phi$ as a monotonically
increasing function of time,

\be \phi-\phi_{0}=t.\ee So the Potential and the kinetic energy of
the scalar field take the following forms respectively,

\be V=\frac{1}{\kappa^2}\left[\frac{3A^2 f^2}{ (\phi-\phi_{0})^{2(1-f)}}-\frac{A f(1-f)}{
(\phi-\phi_{0})^{(2-f)}}\right]-
\frac{(1-w_{B}){\rho_{B}}^{(0)}}{2[a_{0}\exp{\{A(\phi-\phi_{0})^f\}}]^{3(1+w_{B})}} ,\ee and,

\be g\dot\phi^2=G(\phi)=\frac{A f(1-f)}{\kappa^2 (\phi-\phi_{0})^{(2-f)}}-
\frac{(1+w_{B}){\rho_{B}}^{(0)}}{2[a_{0}\exp{\{A(\phi-\phi_{0})^f\}}]^{3(1+w_{B})}}.\ee The energy density and
the pressure of the scalar field are expressed as,

\be \rho_{\phi}=\frac{3A^2 f^2}{\kappa^2 (\phi-\phi_{0})^{2(1-f)}}
-\frac{{\rho_{B}}^{(0)}}{[a_{0}\exp{\{A(\phi-\phi_{0})^f\}}]^{3(1+w_{B})}},\ee and

\be p_{\phi}=\frac{1}{\kappa^2}\left[\frac{2A f(1-f)}{ (\phi-\phi_{0})^{(2-f)}}-\frac{3A^2 f^2}{
(\phi-\phi_{0})^{2(1-f)}}\right]-
\frac{w_{B}{\rho_{B}}^{(0)}}{[a_{0}\exp{{A(\phi-\phi_{0}})^f\}}]^{3(1+w_{B})}}.\ee

The scale factor admits scaling solution as already noticed and
the scaling of $\rho_{\phi}$ becomes sloth as $V(\phi)$ starts
dominating over the kinetic energy. However, synchronized scaling
of $\rho_{\phi}$ and $p_{\phi}$ is not enough, realistic tracking
behaviour is necessary to solve the coincidence problem. Thus, it
remains to be checked if the scalar field is a tracker field.

\subsection{Presence of background Radiation}
In the radiation dominated era $w_{B}=w_{r}=1/3$, the state
parameter of the scalar field is given by,

\be w_{\phi}= \frac{3a_{0}^4e^{4A(\phi-\phi_{0})^f}[2Af(1-f)-3A^2f^2(\phi-\phi_{0})^f]-
\kappa^2{\rho_{r}}^{(0)}(\phi-\phi_{0})^{(2-f)}} {3A^2f^2a_{0}^4e^{4A(\phi-\phi_{0})^f}(\phi-\phi_{0})^f
-\kappa^2{\rho_{r}}^{(0)}(\phi-\phi_{0})^{(2-f)}},\ee
where, $\rho_{r}$ stands for the energy density of
radiation. Now, $w_{\phi}\geq 1/3$ requires,

\be3Afa_{0}^4[5Af(\phi-\phi_{0})^f-3(1-f]e^{4A(\phi-\phi_{0})^f}
+\kappa^2{\rho_{r}}^{(0)}(\phi-\phi_{0})^{(2-f)}\leq 0.\ee
The above condition does not hold since the first
term is positive and contributes dominantly for a monotonically increasing function of $\phi$. So, $w_{\phi}<
1/3$. Thus we conclude that $w_{\phi}<w_{r}$. The form of the potential (36) is

\be V=\frac{1}{\kappa^2}[\frac{3A^2 f^2}{ (\phi-\phi_{0})^{2(1-f)}}-\frac{A f(1-f)}{ (\phi-\phi_{0})^{(2-f)}}]-
\frac{{\rho_{r}}^{(0)}}{3a_{0}^4~e^{4A(\phi-\phi_{0})^f}}.\ee

The above form of the potential is a tracker field, since
$\Gamma>3/2$, \cite{tc} requires,

\[\frac{Af^2(1-f)}{\kappa^4}[18A^2f^2(\phi-\phi_{0})^{2f}
-3Af(3-f)(\phi-\phi_{0})^f+\frac{(1-f)(2-f)}{2}]e^{8A(\phi-\phi_{0})^f}\]
\[-\frac{{\rho_{r}}^{(0)}}{3\kappa^2a_{0}^4}[48A^3f^3(\phi-\phi_{0})^{2(1+f)}
- 76A^2f^2(1-f)(\phi-\phi_{0})^{(2+f)}
 +2Af(1-f)(19-10f)(\phi-\phi_{0})^2-\]
\[(1-f)(2-f)(3-f)(\phi-\phi_{0})^{(2-f)}]e^{4A(\phi-\phi_{0})^f}
+\frac{4{{\rho_{r}}^{(0)}}^2}{9a_{0}^8}[2Af(\phi-\phi_{0})^4-(1-f)(\phi-\phi_{0})^{(4-f)} ]>0,\]which is obvious
because, with evolution $\phi$ grows and the very first term of the above equation, which is positive definite,
is most dominant. Straight forward, but lousy calculation also shows that
$|\frac{\Gamma'}{\Gamma(\frac{V'}{V})}|\ll 1$, ie., $\Gamma$ remains almost constant. Hence the potential (42)
is indeed a tracker field.

\subsection{Presence of Baryonic and non-Baryonic matter}
During the present matter dominated era, $w_{B}=w_{m}=0$, the
state parameter of the scalar field has the following expression,

\be w_{\phi}=\frac{2Af(1-f)-3A^2f^2(\phi-\phi_{0})^f}{3A^2f^2(\phi-\phi_{0})^f-
\kappa^2\rho_{m}^{(0)}a_{0}^{-3}(\phi-\phi_{0})^{2-f}e^{-3A(\phi-\phi_{0})^f}}\ee
The above form of the state
parameter is always negative, since, it requires, $(\phi-\phi_{0})^f>\frac{2(1-f)}{3Af}$, which is true since
$\phi$ has been chosen to increase monotonically with time. Thus, we arrive at the fact that $w_{\phi}<w_{B}$.
The potential (36) in the matter dominated era takes the following form,

\be V=\frac{1}{\kappa^2}[\frac{3A^2 f^2}{ (\phi-\phi_{0})^{2(1-f)}}-\frac{A f(1-f)}{ (\phi-\phi_{0})^{(2-f)}}]-
\frac{{\rho_{m}}^{(0)}}{2a_{0}^3~e^{3A(\phi-\phi_{0})^f}} .\ee
This potential is again a tracker field, since
$\Gamma>3/2$, \cite{tc} now requires

\[\frac{Af^2(1-f)}{\kappa^4}[18A^2f^2(\phi-\phi_{0})^{2f}
-3Af(3-f)(\phi-\phi_{0})^f+\frac{(1-f)(2-f)}{2}]e^{6A(\phi-\phi_{0})^f}\]
\[-\frac{{\rho_{m}}^{(0)}}{2\kappa^2a_{0}^3}[27A^3f^3(\phi-\phi_{0})^{2(1+f)}
- 54A^2f^2(1-f)(\phi-\phi_{0})^{(2+f)}
 +3Af(1-f)(11-6f)(\phi-\phi_{0})^2-\]
\[(1-f)(2-f)(3-f)(\phi-\phi_{0})^{(2-f)}]e^{3A(\phi-\phi_{0})^f}
+\frac{3{{\rho_{m}}^{(0)}}^2}{8a_{0}^6}[3Af(\phi-\phi_{0})^4-2(1-f)(\phi-\phi_{0})^{(4-f)}]>0,\]
which is obvious
as before. Straight forward calculation here again shows that $|\frac{\Gamma'}{\Gamma(\frac{V'}{V})}|\ll 1$,
ie., $\Gamma$ remains almost constant. Hence the potential (44) is indeed a tracker field. Thus, our fourth and
final assertion is proved.
\subsection{Presence of both background radiation and the matter}
In the presence of both the radiation and matter in the form of
dust, we have $w_{\phi}<w_{B}$. The potential has the following
form,

\be V=\frac{1}{\kappa^2}[\frac{3A^2 f^2}{ (\phi-\phi_{0})^{2(1-f)}}-\frac{A f(1-f)}{ (\phi-\phi_{0})^{(2-f)}}]-
\frac{{\rho_{r}}^{(0)}}{3a_{0}^4~e^{4A(\phi-\phi_{0})^f}}-
\frac{{\rho_{m}}^{(0)}}{2a_{0}^3~e^{3A(\phi-\phi_{0})^f}},\ee
which contains algebraic sum of two inverse powers
and two inverse exponents. The condition $\Gamma>3/2$ requires,

\[\frac{Af^2(1-f)}{\kappa^4}[18A^2f^2(\phi-\phi_{0})^{2f}-3Af(3-f)(\phi-\phi_{0})^f
+\frac{(1-f)(2-f)}{2}]e^{8A(\phi-\phi_{0})^f}\] \[+~ (terms~which~
vanish ~as ~\phi ~ is ~large) >0. \]

Clearly the above condition is satisfied and thus the potential is
a tracker field.

\section{Concluding remarks}
The cosmological solution in the form $a = a_{0}e^{(At^f)}$ with $a_{0}>0, A>0$ and $0<f<1$ should be treated as
dark energy model leading to late time cosmic acceleration, rather than intermediate inflation at early
Universe. In the context of general theory of relativity with a minimally coupled scalar field, the accompanying
potential in the form of double inverse power, is a tracker field ($\Gamma>1$ and is slowly varying). Thus the
coincidence problem gets solved. This has been shown in section 3.1. Even a minimally coupled theory with
non-canonical kinetic energy reproduces the same set of solutions. The potential is also in the same form of
double inverse power, which again satisfies the tracker condition ($\Gamma>3/2$ and is slowly varying). This has
been revealed in section 3.2. However, such solution of the gravitational action with non-canonical kinetic
energy does not always carry a potential which is tracker. A non-tracking potential in the form of double
inverse exponents has been studied in section 3.3. Such solution $a = a_{0}e^{(At^f)}$ is also admissible in the
presence of background matter. The potential here is in the form of the sum of double inverse power and an
inverse exponent, when the background is either radiation or dust. In the presence of both, the potential has a
form of the sum of double inverse power and double inverse exponents. These completely new type of potentials
are found to be tracker fields since, for $w_{\phi}<w_{B},\Gamma>3/2$ and is slowly varying. It is noticed that
potential in the form of double inverse power is a tracker field while that with double inverse exponents is
not. However a combination of two is again tracking. This is obvious, since as $\phi$ increases inverse
exponents become less dominant. Finally, the equation of state $w_{\phi}\rightarrow -1$ and the solution in the
above form $a = a_{0}e^{(At^f)}$, asymptotically behaves as
de-Sitter solution. \\
Thus, we conclude that a minimally coupled scalar field admitting
the above form of solution of the scale factor $a =
a_{0}e^{(At^f)}$, has all the nice features to account for the
dark energy of the present Universe.

\end{document}